\documentclass[12pt]{iopart}
\usepackage{cite}
\usepackage{graphicx}
\usepackage{amssymb}
\def\9{\,\rmd}
\def\erf{\mathop{\mathrm{erf}}}
\def\E{\mathrm{E}}

\begin{document}

\title[Roughness bias originating from background subtraction]
      {Estimation of roughness measurement bias originating from background subtraction}

\author{D Ne\v{c}as}

\address{RG Plasma Technologies, CEITEC, Masaryk University, Kamenice 5, 625 00 Brno, Czech Republic}
\address{CEITEC, Brno University of Technology, Purky\v{n}ova 123, 612 00 Brno, Czech Republic}
\ead{yeti@physics.muni.cz}

\author{P Klapetek}

\address{Czech Metrology Institute, Okru\v{z}n\'{\i} 31, 638 00 Brno, Czech Republic}
\ead{pklapetek@cmi.cz}

\author{M Valtr}

\address{Czech Metrology Institute, Okru\v{z}n\'{\i} 31, 638 00 Brno, Czech Republic}
\address{CEITEC, Brno University of Technology, Purky\v{n}ova 123, 612 00 Brno, Czech Republic}
\ead{mvaltr@cmi.cz}

\begin{abstract}
When measuring the roughness of rough surfaces, the limited sizes of scanned
areas lead to its systematic underestimation.  Levelling by polynomials and
other filtering used in real-world processing of atomic force microscopy data
increases this bias considerably. Here a framework is developed providing
explicit expressions for the bias of squared mean square roughness in the case
of levelling by fitting a model background function using linear least
squares.  The framework is then applied to polynomial levelling, for both
one-dimensional and two-dimensional data processing, and basic models of
surface autocorrelation function, Gaussian and exponential.  Several other
common scenarios are covered as well, including median levelling, intermediate
Gaussian--exponential autocorrelation model and frequency space filtering.
Application of the results to other quantities, such as Rq, Sq, Ra and~Sa is
discussed.  The results are summarized in overview plots covering a range of
autocorrelation functions and polynomial degrees, which allow graphical
estimation of the bias.
\end{abstract}

%\verb|$Id: article2.tex 11587 2019-12-31 21:02:46Z yeti $|

\vspace{2pc}
\maketitle

\section{Introduction}

Surface roughness is a ubiquitous phenomenon which influences many
interactions of an object with outer
world---mechanical~\cite{He19,Alipour19,Pravinraj18},
optical~\cite{Trost18,Macias16,Tan16},
chemical~\cite{Shen19},
biological~\cite{Kocer17}, and others.
Its influence is particularly large in the nanoscience and nanotechnology
fields, where object sizes are comparable to characteristic dimensions of
roughness (height and/or lateral) which arise naturally during deposition and
processing of materials.  Whether roughness is considered a defect to be
minimized or potentially useful property to be optimized, it must be measured.
Scanning Probe Microscopy (SPM) techniques, such as Atomic Force Microscopy
(AFM), allow direct measurement of nanoscale roughness---while optical
techniques allow its characterisation in the frequency domain~\cite{Zhao00}.
Larger-scale roughness can be measured by profilometry techniques, of which
mechanical profilometry in some sense analogous to SPM.

Two approaches to measurement should be distinguished.  In an industrial
context reproducibility is key and thus the focus is on procedures and
parameters defined by
standards~\cite{ISO-4287,ASME-B46.1,ISO-25178,ISO-19606}.  It may be of less
concern if these parameters are those occurring in theoretical models or if
they correspond to parameters of a hypothetical random process.  On the other
hand, in basic research the instruments, methods and samples are frequently all
non-standard.  Simultaneously, it is important to estimate parameters that
correspond to theoretical descriptions.  This can be either because they are
themselves interesting, for instance in determining the universality class
for a growth mechanism~\cite{Barabasi95,Zhao00}. Or they appear in physical
theories describing interactions with rough surfaces. Probably the most
interesting parameter is squared mean square roughness $\sigma^2$ which
directly appears in optics---together with similar quadratic quantities such
as spectral densities of spatial frequencies and (cross)correlation
functions~\cite{Trost18,Necas14,Cermak18}.  Here we will approach roughness
from this second standpoint.

Surface roughness is never measured using data from an infinitely large region
with infinite resolution.  The resolution is always finite---in contact
scanning methods (AFM or profilometry) limited by finite probe size, in
optical methods by finite wavelength.  The measurement area is also always
finite and seldom even encompasses the entire sample, in particular in direct
measurements.  In fact, in AFM we regularly measure tiny fractions of the
surface---and instead of rigorous statistical justification for
representativeness of the results we just have hope that no evil forces
conspired to plant non-representative surface regions under the probe.

Still, conceptually, the statistical character of roughness parameters is
acknowledged~\cite{Zhao00,Klapetek18}.  We imagine an infinite ensemble of
surfaces (possibly infinite themselves), usually corresponding formally to
a random process, which may or may not be wide-sense stationary.  Measurement
of non-stationary fractal surfaces in an interval of scales in which they do
exhibit self-affinity adds its own set of difficulties~\cite{Zhao00,Brown18}.
Here we will focus on roughness generated by stationary processes---and
estimation of their parameters using a finite measurement of one realization.
In particular, we will study the consequences of finite measurement area and
levelling/background subtraction.

One obvious consequence is that the estimated parameter, for example mean
square roughness $\sigma$, estimated from a profile of length $L$ by
\begin{equation}
\hat\sigma^2 = \frac1L \int_0^L z(x)^2 \9x
\label{sigma-simple}
\end{equation}
is itself a random variable, as we denote with a hat.  It has a dispersion,
which is possibly large~\cite{Zhao00}.  Definition~\eref{sigma-simple}
corresponds to mean square roughness Rq for
profiles~\cite{ISO-4287,ASME-B46.1,ISO-19606} and Sq for
images~\cite{ISO-25178-2} defined by roughness measurement standards, with the
subtle conceptual difference discussed above.

The estimate is also biased.  Heights $z(x)$ entering~\eref{sigma-simple} are
levelled to have zero mean value.  This alone introduces
bias, which is well known and discussed for correlated data in classical
signal processing textbooks~\cite{Anderson71,Zhao00,Krishnan15,Box2015}.
However, subtraction of the mean value is rarely the only preprocessing
applied to topographical data before roughness evaluation.  Often local
defects are removed first, although this may be unnecessary as evaluation
algorithms for irregular regions allow excluding arbitrary image parts from
processing~\cite{Necas13}.  Almost universally the mean plane is subtracted to
correct tilt---and frequently not just a plane but a higher order polynomial
to correct scanner bow (or sample warping)~\cite{Klapetek18}.  Misaligned scan
lines need to be aligned for 2D processing, although not necessarily for
line-by-line evaluation. Furthermore, any of specific form removal methods can
be utilized, from frequency-space filtering to wavelet processing to
subtraction of specific geometrical shapes such as sphere.

Some of these steps remove background arising from measurement imperfections
(scanner bow), some remove real base shape of the measured object.  Often they
remove both to some degree---and by intentionally removing certain degrees of
freedom they also always remove inadvertently a part of the roughness. For
instance in the case of the mean value, we subtract it because the measured
surface height $h(x)$ is not the roughness signal $z(x)$.  It is offset by
some background, in this case a constant base height $B$:
\begin{equation}
h(x) = z(x) + B\;.
\end{equation}
The background $B$ is non-random (at least from the roughness standpoint), but
unknown.  We estimate it as the mean value of the heights
\begin{equation}
\hat B = \frac1L \int_0^L h(x) \9x
\label{B-simple}
\end{equation}
because the expected value of $h$ is $\E[h]=B$.  However, the subtraction
of $\hat B$ instead of true $B$ removes not just $B$ but also a part of the
roughness.  Although the expected value of $z$ is zero, the mean value of
$z(x)$ over a finite interval is a random variable, not zero---yet we make it
zero anyway.

\begin{figure}
\includegraphics{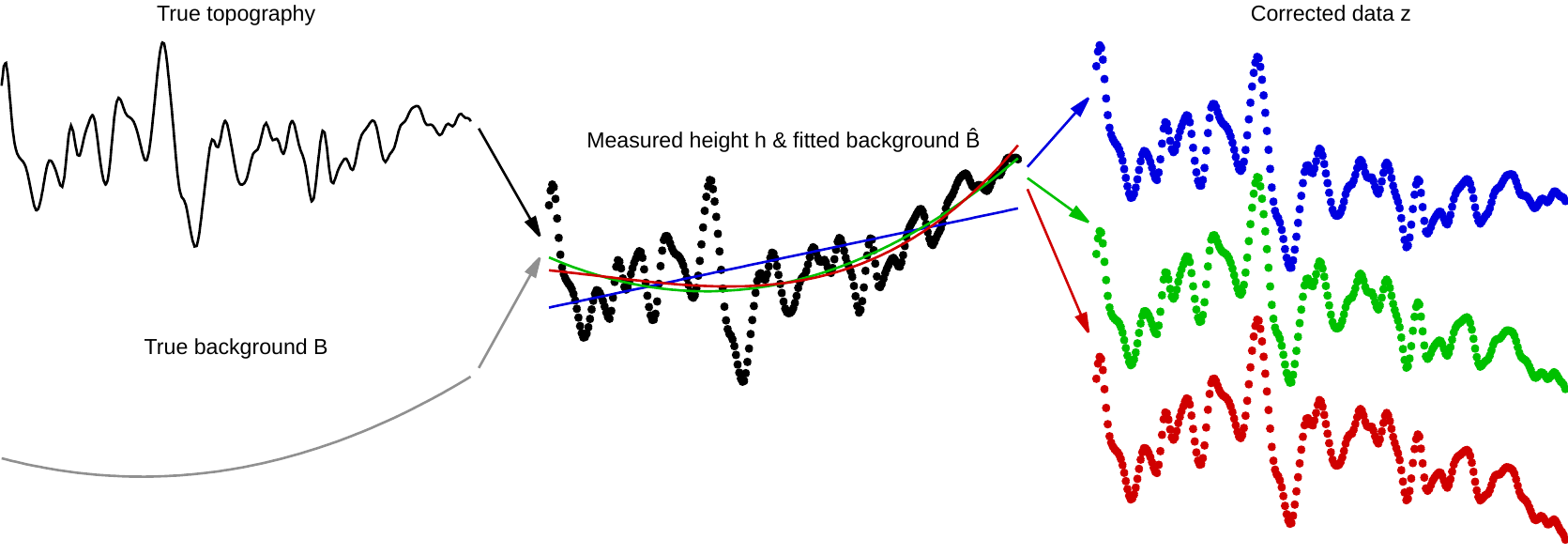}
\caption{General scheme of data distortion by background removal. The true
topography and true background (here bow) combine in the measured data.
The exact background type is not known and must be chosen from a set of
models fitted to the data, here polynomials of degrees 1, 2 and~3.  The
fitting is greedy and subtracts not just the true background, but also
roughness components which randomly match it.  The corrected data are then
missing these components.}
\label{fig:dof-removal}
\end{figure}

This is illustrated in \eref{fig:dof-removal} for a second-order
polynomial~$B$.  A second-order polynomial background was added to an `ideal'
rough signal.  Then a polynomial background was fitted and removed.  We seldom
know the exact background type and each choice levels the surface differently.
Furthermore, even if the correct degree is chosen, the levelled surface
differs from the original ideal one.

As already noted above, real AFM or profilometry data are sets of discrete
values $z_k$, not continuous functions $z(x)$.  Integrals such as
\eref{sigma-simple} or~\eref{B-simple} are approximated by summations, for
instance
\begin{equation}
\hat\sigma^2 = \frac1N \sum_{k=1}^N z_k^2\;.
\label{mu-sum}
\end{equation}
There is a certain arbitrariness in the correspondence between the region
$[0,L]$ and the set of points where heights $z_k$ were measured.  If we state
that $z_k$ were obtained in the centres of sampling steps (or pixel centres
for images), the `measured region' covers $z_k$ and extends a further
half-step to each side.
% Some more classical reference?  Abramowitz?
Formula \eref{mu-sum} then becomes the midpoint
quadrature rule~\cite{Zwilinger92} with only second-order error and
approximates well the integral as long as the sampling step $\Delta=L/N$ is
small compared to the autocorrelation length $\Delta\ll T$.
Since this work focuses on the effect of finite measurement area, i.e.\ loss
of low-frequency information, we will not dwell on the loss of high-frequency
information and will assume the sampling is sufficiently fine. Therefore,
continuous functions will be used in the following analysis (instead of sets
of discrete sampled values $z_k$).

%Requiring the that surface is not undersampled results in an inherent
%speed-area trade-off, illustrated in~\fref{fig:profile-sampling}.  Sampling
%the same region more finely (\fref{fig:profile-sampling}a) is usually easier to
%achieve because the probe speed does not need to increase.  However, when the
%sampling step is much finer than the surface autocorrelation length we measure
%much more data, but add very little new information because the values are
%highly correlated.  \Fref{fig:profile-sampling}b illustrates the case when the
%number of measurements is the limiting factor, not probe speed.  Again, there
%is a trade-off is between measuring uncorrelated data and losing spatial
%information (too coarse step) and measuring highly correlated data and wasting
%time (too fine step).  Usually, we need to deal with a combination both
%limitations. {\it XXX And someone must have discussed this in detail and
%developed some optimal sampling models.  Cite!}
%
%\begin{figure}
%\includegraphics{profile-sampling.pdf}
%\caption{Sampling illustration for 1D data.
%(a) Three sampling densities of the same profile:
%undersampled, almost uncorrelated values with no spatial information; sampled
%with $\Delta\approx 0.7T$; and very finely sampled.
%(b) The same densities displayed with the same number samples---a condition
%more closely matching limitations in AFM measurements.}
%\label{fig:profile-sampling}
%\end{figure}

\section{Mean value subtraction}
\label{sect:mvs}

The root mean square roughness $\sigma$ is estimated from heights $z$ in
a finite-size region $[0,L]$
\begin{equation}
\hat\sigma^2 = \frac1L \int_0^L [z(x) - \hat\mu]^2 \9x\;,
\quad\hbox{where}\quad
\hat\mu = \frac1L \int_0^L z(x) \9x\;.
\label{sigma-integral}
\end{equation}
For simplicity, we will consider one-dimensional (1D) data here.

The formula for $\hat\sigma^2$ is now always presented with explicit
subtraction of of $\hat\mu$.  Instead we simply say that `the mean value of
heights is zero'.  However, this confounds two distinct statements:
\begin{itemize}
\item the expected value of roughness signal $z$ is zero $\E[z]=0$, and
\item the mean value of measured data is made zero by preprocessing.
\end{itemize}
The second is the source of bias since $\hat\mu$ is a random variable, not
identically equal to zero even when its expected value is: $\E[\hat\mu]=0$.

We now briefly reproduce the classical result for the bias caused by mean
value subtraction~\cite{Anderson71,Zhao00,Krishnan15,Box2015}.  The derivation
provides an outline for how the more complex cases will be treated in
\sref{sect:realbg}. Expanding the square in \eref{sigma-integral} gives
\begin{equation}
\hat\sigma^2 = \frac1L \int_0^L z^2(x) \9x - \hat\mu^2\;.
\label{sigma-expanded}
\end{equation}
We would like to know the expected value of the estimate $\E[\hat\sigma^2]$.
The expected value of the first term on the right hand side of
\eref{sigma-expanded} is $\sigma^2$. Hence the second term gives the
bias---which is always negative. Writing
\begin{equation}
\hat\mu^2
    = \frac1L \int_0^L z(x) \9x
      \times \frac1L \int_0^L z(x') \9x'
    = \frac1{L^2} \int_0^L \int_0^L z(x) z(x') \9x' \9x
    \label{mu-product}
\end{equation}
and using coordinate transformation $(x,x')=(v,u+v)$ we obtain
% we also split the result into positive and negative u part and transform...
\begin{equation}
\hat\mu^2 = \frac1{L^2} \int_0^L \int_0^{L-u} z(v) z(v+u) \9v \9u
            + \frac1{L^2} \int_0^L \int_u^L z(v) z(v-u) \9v \9u\;.
\label{mu-split}
\end{equation}
Expected value calculation can be interchanged with integration.  The
expected value of either integrand is the autocorrelation function (ACF) of
the signal
\begin{equation}
G(t) = E\left[z(x)z(x+t)\right] = E\left[z(x)z(x-t)\right]\;.
\end{equation}
Note that $G(0)=\E[z(x)^2]=\sigma^2$. Substituting this result into
\eref{sigma-expanded} gives the classic final
expression~\cite{Anderson71,Krishnan15}
% Anderson, The Statistical Analysis of Time Series, Wiley (1971), ISBN
% 0-471-04745-7, p.448, Equation 51. -- discrete formulation, probably
%
% These are also discrete versions:
% Law and Kelton, p.285. This equation can be derived from Theorem 8.2.3 of
% Anderson. It also appears in Box, Jenkins, Reinsel, Time Series Analysis:
% Forecasting and Control, 4th Ed. Wiley (2008), ISBN 978-0-470-27284-8, p.31.
%
% Who shows exactly our continuous version?
\begin{equation}
\E[\hat\sigma^2]
    = \sigma^2 - \frac2L \int_0^L \left(1-\frac{t}{L}\right) G(t)\9t
    = \sigma^2 - 2 \int_0^1 (1-t) G(Lt) \9t
\label{mu-classic}
\end{equation}
Since the bias is always negative and proportional to $\sigma^2$, it is
convenient to introduce the relative bias $\beta$
\begin{equation}
\E[\hat\sigma^2] = \sigma^2 (1 - \beta)
\end{equation}
to simplify notation.  If we know $\beta$, replacing $\hat\sigma^2$ with
$\hat\sigma^2/(1-\beta)$ corrects the bias.

In order to see how the bias typically behaves, we evaluate it for a simple
prabolic model of ACF $G(t)=\sigma^2(1-t^2/T^2)$ for $t<T$ and zero otherwise.
Then for $L\ge T$
\begin{equation}
\beta_\mathrm{1D}
    = \frac43 \alpha \left(1 - \frac38 \alpha\right)
    \approx \frac 43 \alpha
\quad\hbox{and}\quad
\beta_\mathrm{2D}
    = \pi \alpha^2 \left(1 - \frac4{15} \alpha\right)
    \approx \pi \alpha^2
\end{equation}
in one and two dimensions, respectively.
The approximations hold for $\alpha=T/L$ small.  The numerical factors such
as $4/3$ and $\pi$ change somewhat with the exact form of the ACF, but
generally are of the same order of magnitude.

The important point is that the relative systematic error of $\hat\sigma^2$
due to finite-area bias behaves approximately as $\alpha$ and $\alpha^2$ for
1D and 2D data, respectively.  More generally, it behaves like $\alpha^D$
where $D$ denotes the dimension~\cite{Zhao00}.  It does \emph{not} depend on
the number of measured values $N$ (provided it is sufficiently large).
Increasing $N$ without making the measurement area larger is of no help and
Bessel's correction
\begin{equation}
\hat\sigma^2 = \frac1{N-1} \sum_{k=1}^N z_k^2\;,
\end{equation}
which replaces $N$ with $N-1$ is ineffective.  For correlated data the
correction factor is not $1-1/N$, but akin to $1-c\alpha$ or $1-c\alpha^2$,
where $c$ is some constant of order of unity.

Almost all roughness measurements involve correlated data.  If we measure with
such a large sampling step that the height values are uncorrelated we lose all
spatial information about the roughness.  This is rarely desirable---and also
rarely possible, since in scanning methods the feedback loop then cannot keep
up with the surface topography, whereas in optical methods this usually means
averaging too large regions of the surface in one pixel.

\section{Real-world background subtraction}
\label{sect:realbg}

Background subtraction methods used for SPM are much more complex than mere
mean value subtraction~\cite{Schouterden96,Fogarty06,Gimeno15,Klapetek18}.
In order to evaluate roughness correctly, we must take into account which
degrees of freedom or spatial frequencies would contribute to the desired
result, but were removed by preprocessing.  This is not trivial to start with
and certainly not helped by AFM data processing software, which can apply
plane levelling or row alignment automatically, possibly without the user even
noticing (depending on the software and settings).
% It seems SPIP applies plane levelling and WsXM aligns rows, according
% to the confusing e-mail exchange with Manuela Bras, but that may be just some
% default settings.  We should not call names unless verified...
And, of course, no AFM software currently attempts to estimate the resulting
bias.

It is common to process AFM image data row by row because roughness properties
can often be determined more reliably in the direction of the fast scanning
axis~\cite{Dumas93,Zhao00,Necas13,Klapetek18}.  This means that levelling is
applied to individual image rows instead of (or in addition to) the entire
image.  The operation of mutual alignment of scan lines is colloquially
referred to as `flatten'~\cite{Schouterden96,Fogarty06,Gimeno15}.  However, we
explicitly call it scan line correction for clarity.

Results for individual rows then may be summed or averaged.  The result
for each row is biased as if we processed 1D data, not 2D. The same holds if
any row-wise preprocessing is applied, such as removal of mean value from each
individual row.  It is, therefore, quite rare that the bias corresponds to the
2D case, even for image data.

\subsection{Linear-fit background}

Removal of tilt, bow or higher order polynomial backgrounds has two basic
steps. Fitting a background function $B(x)$ to the data using the linear least
squares method, and subtraction of the fitted (estimated) background
$\hat B(x)$.  This section presents a general framework for evaluating bias
resulting from background subtraction by linear fitting. Not all background
removal methods are linear---for instance the subtraction of median or
a fitted spherical surface is non-linear, but most common ones are.

A linear fitting function satisfies
\begin{equation}
B(x) = \sum_j a_j \varphi_j(x)
     = \sum_j a_j \frac{\partial B(x)}{\partial a_j}\;,
\label{B-is-linear}
\end{equation}
where $\varphi_j$ are basis functions (for instance powers of $x$) and $a_j$
the corresponding coefficients---fitting parameters. The fit minimises the
residual sum of squares, therefore
\begin{equation}
\frac{\partial}{\partial a_j} \int_0^L [z(x) - \hat B(x)]^2 \9x = 0\;.
\end{equation}
These two relations allow expanding the expression for $\hat\sigma^2$ as
follows:
\begin{equation}
\hat\sigma^2
    = \frac1L \int_0^L [z(x) - \hat B(x)]^2 \9x
    = \frac1L \int_0^L z(x)^2 \9x - \frac1L \int_0^L \hat B(x)^2 \9x\;.
    \label{sigma-B}
\end{equation}
Again, the second term gives the bias.

The linear fit corresponds to an orthogonal projection onto a linear function
subspace spanning $\varphi_j$.  It can, therefore, be assumed without loss of
generality that $\varphi_j$ are orthonormal---and we will do so in order to
simplify notation.  Some sets of $\varphi_j$ naturally come as orthonormal,
for instance sines and cosines in frequency-space filtering, and this holds
also for some wavelet bases.  If required, any set of linearly independent
basis functions can be made orthonormal by an orthogonalization process such
as Gram--Schmidt, followed by normalization.  In the case of polynomial
backgrounds, orthonormal polynomials can be directly chosen as the
basis~$\varphi_j$.

For orthonormal $\varphi_j$ the estimated coefficients are simple scalar
products
\begin{equation}
\hat a_j = \int_0^L z(x) \varphi_j(x) \9x
\label{a_j}
\end{equation}
and thus
\begin{equation}
\int_0^L \hat B(x)^2 \9x
    = \sum_j \hat a_j^2
    = \sum_j \int_0^L z(x) \varphi_j(x) \9x
             \int_0^L z(x') \varphi_j(x') \9x'\;.
\end{equation}
Note that these $\hat a_j$ are not the best estimators of the
coefficients---the problem dual to ours, i.e.\ linear fitting of correlated
data, has a more complex solution~\cite{Lee04}.  However, \eref{a_j}
corresponds to levelling methods used in practice.

Transforming this expression in the same manner as~\eref{mu-product} results
in
\begin{equation}
\int_0^L \hat B(x)^2 \9x
    = \sum_j \int_0^L \int_0^{L-u}
                          z(v) z(v+u) \varphi_j(v) \varphi_j(v+u) \9v \9u\;.
\end{equation}
In calculation of the expected value we note that $\varphi_j$ are not
realizations of a random process and can be factored out
\begin{equation}
\E[z(v) z(v+u) \varphi_j(v) \varphi_j(v+u)]
    =  \E[z(v) z(v+u)] \varphi_j(v) \varphi_j(v+u)\;,
\end{equation}
which after some rearrangement gives the final expression
\begin{equation}
\E[\hat\sigma^2] = \sigma^2 - 2^D \int_0^1 G(tL) C(t) \9t \;.
\label{sigma-C}
\end{equation}
In each dimension we sum two integrals in \eref{mu-split}, so each gives
factor 2.  In $D$ dimensions intervals and integrals are $D$-dimensional,
interval $[0,1]$ stands for $[0,1]^D$, etc. Function $C$ is determined
entirely by the set of orthonormal basis functions and the interval
\begin{equation}
C(t) = \sum_j \int_0^{L(1-t)} \varphi_j(v) \varphi_j(v+Lt) \9v
     = \sum_j c_j(t)\;.
\label{C}
\end{equation}
By considering the single constant basis function $\varphi_0=1/\sqrt{L}$ we
recover mean value subtraction formulae from section~\ref{sect:mvs}.

Since roughness is evaluated under the assumption `mean value of $z$ is zero',
the basis always includes the constant function.  If we subtract some other
type of background, there are two possibilities.  Either this already ensures
zero mean value and then the linear span indeed includes constant functions.
Or it does not and we must subtract the mean value afterwards to make it zero.
However, the constant function is then independent and can be simply added to
the basis, merging the two setps.

\subsection{Autocorrelation of a linear function space}

Function $C(t)$ is a curious characteristic of the background removal method.
Although it is evaluated using a concrete orthornomal basis $\varphi_j$ in
\eref{C}, it does not depend on the choice of the basis---this can be
easily seen if we express $\varphi_j$ in a different basis.
The function describes the subtraction of projection onto the linear function
space spanned by $\varphi_j$.  For instance, it is immaterial whether we
actually fit orthonormal polynomials or just plain powers $x^j$ during
background subtraction because their linear span is the same.

In this manner $C(t)$ characterizes the correlations in an entire linear
subspace of functions.  On an infinite interval it has perhaps a clearer
interpretation.  In such case we can express $\varphi_j$ using the Fourier
transform
\begin{equation}
\varphi_j(v) = \int_{-\infty}^\infty \exp(-2\pi\rmi \xi v) \Phi_j(\xi) \9\xi\;.
\end{equation}
Subtituting it into formula \eref{C} gives according to the correlation
theorem~\cite{Bracewell99}
\begin{equation}
C(t) = \sum_j
       \int_{\infty}^\infty \exp(-2\pi\rmi \xi v) |\Phi_j(\xi)|^2 \9\xi\;,
\end{equation}
where $|\Phi_j|^2$ is the spectral density of $\phi_j$.  Therefore,
$C(t)$ is the Fourier transform of
\begin{equation}
W(\xi) = \sum_j |\Phi_j(\xi)|^2\;,
\end{equation}
which is the total spectral density of the orthonormal basis, i.e.\ in some
sense the spectral density of the linear subspace.
% Someone surely had to define this before?
This is an useful intuition which can be transferred to finite intervals,
even though the formulae from this paragraph do not hold exactly there,
polynomials are not a useful basis on infinite intervals, etc.

\subsection{One-dimensional polynomial background}

\begin{table}
\caption{\label{tab:djpoly} Polynomials $d_j$ corresponding to individual
Legendre polynomials according to \eref{legendre-cjdj} and used to construct
expressions for specific polynomial background removal types.  The argument of
$d_j$ is $x=t^2$.}
\begin{indented}
\item[]\begin{tabular}{@{}llll}
\br
$j$&$d_j(x)$\\
\mr
0&$0$\\
1&$1$\\
2&$2 - 3x$\\
3&$3 - 11x + 10x^2$\\
4&$4 - 26x + 55x^2 - 35x^3$\\
5&$5 - 50x + 181x^2 - 259x^3 + 126x^4$\\
6&$6 - 85x + 461x^2 - 1099x^3 + 1176x^4 - 462x^5$\\
7&$7 - 133x + 1001x^2 - 3499x^3 + 6126x^4 - 5214x^5 + 1716x^6$\\
8&$8 - 196x + 1946x^2 - 9274x^3 + 23451x^4 - 32241x^5 + 22737x^6 - 6435x^7$\\
%9&$9-276x+3486x^2-21594x^3+73501x^4-144353x^5+162877x^6-97955x^7+24310x^8$\\
%10&$10-375x+5862x^2-45618x^3+199627x^4-522731x^5+835549x^6-798083x^7+418132x^8-92378x^9$\\
\br
\end{tabular}
\end{indented}
\end{table}

Orthonormal polynomial basis on the interval $[0,L]$ is formed by shifted and
scaled Legendre polynomials $\mathrm{P}_j$~\cite{Abramowitz64}:
\begin{equation}
\varphi_j(x)
    = \sqrt{\frac{2j+1}L}\,\mathrm{P}_j\left(\frac{2x}L - 1\right)\;.
    \label{Legendre-1D}
\end{equation}
Evaluation of integrals~\eref{C} leads to
\begin{equation}
c_j(t)
    = \left(j + \frac12\right)
      \int_{-1}^{1-2t} \mathrm{P}_j(x) \mathrm{P}_j(x+2t) \9x
    = 1-t - 2t(1-t^2)d_j(t^2)\;,
    \label{legendre-cjdj}
\end{equation}
where functions $d_j$ are listed in \tref{tab:djpoly} for polynomial degree
up to~8 (these and other tedious integrals were evaluated symbolically in
Maxima~\cite{Maxima}).
Polynomials $c_j$ depend on the specific choice of orthonormal basis.
Polynomials $C_j$ which are obtained by summing them up to specific degree
according to~\eref{C} depend only on the linear span covered by the basis.
They are listed for reference in \tref{tab:1d-Cpoly}.

\begin{table}
\caption{\label{tab:1d-Cpoly} Polynomials $C_j$ describing 1D polynomial
background removal of degree $j$.}
\begin{indented}
\item[]\begin{tabular}{@{}llll}
\br
$j$&$C_j(t)$\\
\mr
0&$1 - t$\\
1&$2 - 4t + 2t^3$\\
2&$3 - 9t + 12t^3 - 6t^5$\\
3&$4 - 16t + 40t^3 - 48t^5 + 20t^7$\\
4&$5 - 25t + 100t^3 - 210t^5 + 200t^7 - 70t^9$\\
5&$6 - 36t + 210t^3 - 672t^5 + 1080t^7 - 840t^9 + 252t^{11}$\\
6&$7-49t+392t^3-1764t^5+4200t^7-5390t^9+3528t^{11}-924t^{13}$\\
7&$8-64t+672t^3-4032t^5+13200t^7-24640t^9+26208t^{11}-14784t^{13}+3432t^{15}$\\
8&$9-81t+1080t^3-8316t^5+35640t^7-90090t^9+137592t^{11}-124740t^{13}+61776t^{15}-12870t^{17}$\\
\br
\end{tabular}
\end{indented}
\end{table}

\begin{table}
\caption{\label{tab:1d-gauss} 1D polynomials for Gaussian ACF.}
\begin{indented}
\item[]\begin{tabular}{@{}llll}
\br
$j$&$g_j(\alpha)$&$p_j(\alpha)$\\
\mr
0&$1$&$1$\\
1&$1 - \alpha^2$
    &$2 - \alpha^2$\\
2&$1 + 4\alpha^4$
    &$3 - 4\alpha^2 + 4\alpha^4$\\
3&$1 - \alpha^2 - 6\alpha^4 - 30\alpha^6$
    &$4 - 10\alpha^2 + 24\alpha^4 - 30\alpha^6$\\
4&$1 + 12\alpha^4 + 96\alpha^6 + 336\alpha^8$
    &$5 - 20\alpha^2 + 84\alpha^4 - 240\alpha^6 + 336\alpha^8$\\
5&$1 - \alpha^2 - 16\alpha^4 - 240\alpha^6 - 1680\alpha^8 - 5040\alpha^{10}$
    &$6 - 35\alpha^2 + 224\alpha^4 - 1080\alpha^6 + 3360\alpha^8 - 5040\alpha^{10}$\\
\br
\end{tabular}
\end{indented}
\end{table}

\begin{table}
\caption{\label{tab:1d-exp} 1D polynomials for exponential ACF.}
\begin{indented}
\item[]\begin{tabular}{@{}llll}
\br
$j$&$e_j(\alpha)$&$q_j(\alpha)$\\
\mr
0&$1$&$1 - \alpha$\\
1&$1 + 6\alpha + 6\alpha^2$
    &$1 - 2\alpha + 6\alpha^3$\\
2&$1 + 16\alpha + 96\alpha^2 + 240\alpha^3 + 240\alpha^4$
    &$1 - 3\alpha + 24\alpha^3 - 240\alpha^5$\\
3&$1 + 30\alpha + 390\alpha^2 + 2760\alpha^3 + 11160\alpha^4 + 25200\alpha^5 + 25200\alpha^6$
    &$1 - 4\alpha + 60\alpha^3 - 1440\alpha^5 + 25200\alpha^7$\\
\br
\end{tabular}
\end{indented}
\end{table}

\begin{figure}
\includegraphics{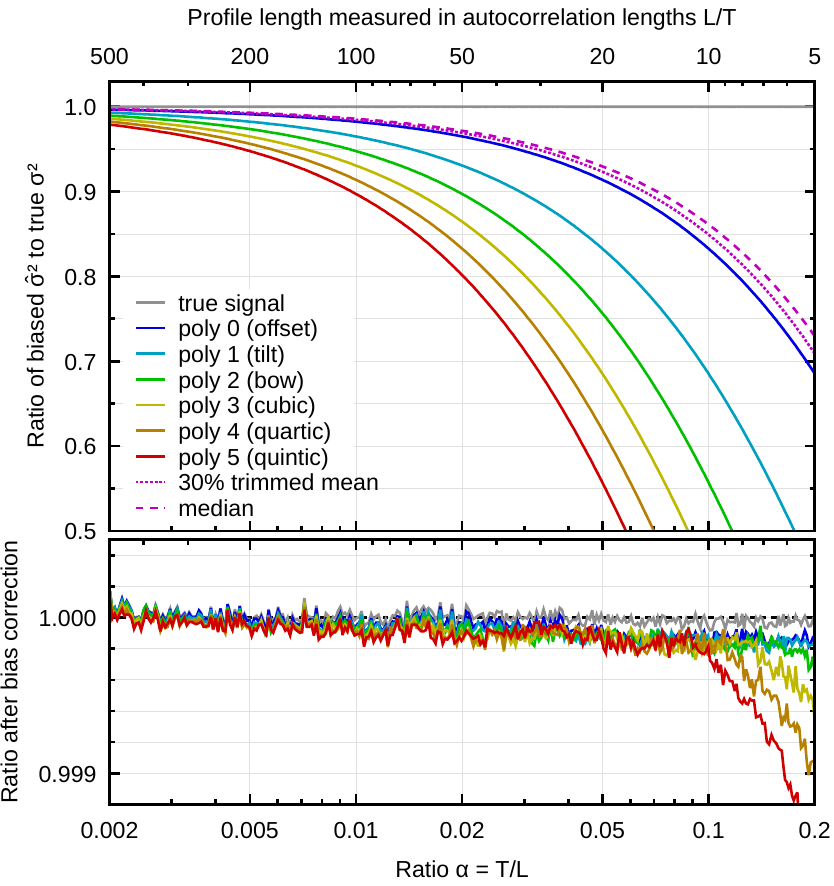}
\caption{Numerical results for polynomial background removal from profiles
with Gaussian ACF:
estimated $\hat\sigma^2$ divided by true $\sigma^2$ for several polynomial
degrees (upper)
and corrected estimate $\hat\sigma^2/(1-\beta)$ (lower).}
\label{fig:stats-1d-gauss}
\end{figure}

In order to obtain concrete expressions for the bias, we still need to specify
the form of the ACF.  Two common models are Gaussian and
exponential~\cite{Zhao00,Klapetek18}
\begin{equation}
G_\mathrm{Gauss}(x) = \sigma^2 \exp(-x^2/T^2)
\quad\hbox{and}\quad
G_\mathrm{exp}(x) = \sigma^2 \exp(-|x|/T)\;.
\label{G-Gauss-exp}
\end{equation}
For $G_\mathrm{Gauss}$ the relative bias resulting from the subtraction of
polynomial degree 0 (i.e.\ mean value) is
\begin{equation}
\beta_0
    = \alpha\sqrt{\pi}\erf(1/\alpha) + \alpha^2 \exp(-1/\alpha^2) - \alpha^2\;.
\end{equation}
More generally
\begin{equation}
\beta_j = (j+1)\left[ \alpha\sqrt{\pi}\erf(1/\alpha)
                      + \alpha^2 \exp(-1/\alpha^2) g_j(\alpha)
                      - \alpha^2 p_j(\alpha)\right]\;,
\end{equation}
where $g_j(\alpha)$ and $p_j(\alpha)$ are polynomials listed in
\tref{tab:1d-gauss}.  The leading-order approximation for small $\alpha$ is
\begin{equation}
\beta_j \sim (j+1)\alpha\left[\sqrt{\pi} - (j+1)\alpha
                              + \frac{j(j+1)(j+2)}6 \alpha^3 \right]\;.
\label{beta-1d-gauss-lo}
\end{equation}
For the exponential ACF we obtain
\begin{equation}
\beta_0 = 2\alpha^2\exp(-1/\alpha) + 2\alpha(1-\alpha)
\end{equation}
and more generally
\begin{equation}
\beta_j = 2(j+1)\left[ \alpha q_j(\alpha)
                       + (-1)^j \alpha^2 \exp(-1/\alpha) e_j(\alpha) \right]\;,
\end{equation}
where $q_j(\alpha)$ and $e_j(\alpha)$ are polynomials listed in
\tref{tab:1d-exp}.  The leading-order approximation for small $\alpha$ is
\begin{equation}
\beta_j \sim 2(j+1)\alpha \left[1 - (j+1)\alpha
                                + j(j+1)(j+2)\alpha^3 \right]\;.
\label{beta-1d-exp-lo}
\end{equation}
As an example, numerical results for the bias of $\hat\sigma^2$ are plotted in
\fref{fig:stats-1d-gauss} for the Gaussian ACF.  The `true' signals
were generated by cutting segments from very long frequency-space synthesized
data.  The figure includes also results for median and 30\% trimmed mean
levelling, which are rather similar to mean value subtraction.

The effectiveness of correcting the estimated $\hat\sigma^2$ by dividing with
$1-\beta$ is evident. There are small residual differences even for the true
signal, stemming from it being still finite, albeit very long, and thus
random. Furthermore, the corrections start to cease being perfect for large
$\alpha$. This is an effect of discretization.

Finally we note that the leading-order approximations \eref{beta-1d-gauss-lo}
and~\eref{beta-1d-exp-lo} would not be changed by putting
$\exp(-1/\alpha^2)=\exp(-1/\alpha)=1-\erf(1/\alpha)=0$.
In fact, since these terms are exponentially small in $1/\alpha$, this change
does not influence any terms in series expansions in powers of $\alpha$ for
$\alpha\to0$.  The same approximation (and conclusion) follows from writing
\begin{equation}
\int_0^1 G(tL) C(t) \9t
    = \int_0^\infty G(tL) C(t) \9t - \int_1^\infty G(tL) C(t) \9t
\label{sigma-C-inf}
\end{equation}
in \eref{sigma-C} and disregarding the second term, exponentially small
compared to the first.  The integral to infinity is generally much easier to
evaluate, in particular in higher dimensions, and allows obtaining formulae
for small $\alpha$~\cite{Zhao00}.

\subsection{Two-dimensional polynomial background}

Two-dimensional orthonormal polynomials on $[0,L_1]\times[0,L_2]$ can be
constructed as separable, i.e.\ products of 1D
polynomials~\eref{Legendre-1D}
\begin{equation}
\varphi_{j_1,j_2}(x_1,x_2) = \varphi_{j_1}(x_1) \varphi_{j_2}(x_2) \;.
\end{equation}
Clearly then
\begin{equation}
\int_0^{L_1} \int_0^{L_2} \varphi_{j_1,j_2}(x_1,x_2)
                          \varphi_{k_1,k_2}(x_1,x_2) \9x_1 \9x_2
    = \delta_{j_1,k_1} \delta_{j_2,k_2}
\end{equation}
and other 2D expressions, such as the integrals in~\eref{C}, reduce to
products of 1D expressions in a similar manner.

%\begin{table}
%\caption{\label{tab:2d-Cpoly} Polynomials $C_j$ describing 2D polynomial
%background removal.  They were divided by the common factor $(1-t_1)(1-t_2)$
%for brevity.}
%\begin{indented}
%\item[]\begin{tabular}{@{}llll}
%\br
%$j$&$C_j(t_1,t_2)/\bigl[(1-t_1)(1-t_2)\bigr]$\\
%\mr
%0&$1$\\
%1&$3 - 2(t_1^2 + t_1 + t_2^2 + t_1)$\\
%2&$6 - 8(t_1^2 + t_1 + t_2^2 + t_2)
%     + 6(t_1^4 + t_1^3 + t_2^4 + t_2^3) + 4t_1t_2(t_1+1)(t_2+1)$\\
%3&$10 - 20(t_1^6 + t_1^5 + t_1^2 + t_1 + t_2^6 + t_2^5 + t_2^2 + t_2)
%    - 12t_1t_2(t_1+1)(t_2+1)(t_1^2 + t_2^2)$\\
% &${} + 20t_1t_2(t_1+1)(t_2+1) + 34(t_1^4 + t_1^3 + t_2^4 + t_2^3)$\\
%\br
%\end{tabular}
%\end{indented}
%\end{table}

Usually the total degree of the polynomial is limited, leading to the
following basis function sets (on square $L_1=L_2=L$):
\begin{itemize}
\item constant $\varphi_0(x_1)\varphi_0(x_2)=1/L$,
\item plane levelling, adding $\varphi_1(x_1)\varphi_0(x_2)$ and
      $\varphi_0(x_1)\varphi_1(x_2)$,
\item quadratic levelling,
      adding $\varphi_2(x_1)\varphi_0(x_2)$, $\varphi_1(x_1)\varphi_1(x_2)$ and
      $\varphi_0(x_1)\varphi_2(x_2)$,
\item cubic levelling, adding the four cubic basis functions,
\item etc.
\end{itemize}
Evaluation of the integral~\eref{C} then results in functions $C(t_1,t_2)$
for 2D polynomial levelling.

However, there are other common choices for the set of polynomials.
Frequently the maximum degrees of $x_1$ and~$x_2$ are chosen separately, in
particular when the image is not square or there are other reasons for using
different levelling along the two axes.  Enumeration of all reasonable
two-dimensional $C(t_1,t_2)$ is not feasible.  Therefore, we instead describe
a procedure for their construction:
\begin{enumerate}
\item Take the set of 2D terms $x_1^{j_1}x_2^{j_2}$ which define the
      polynomial background.
\item For each term look up the corresponding $d_{j_1}$ and $d_{j_2}$ in
      \tref{tab:djpoly}.
\item For each term calculate polynomials $c_{j_1}$ and $c_{j_2}$ according to
      \eref{legendre-cjdj}.  Multiply the two polynomials.
\item Sum the results for all terms.
\end{enumerate}
This procedure is applicable if the set of degrees is convex, i.e.\ if under
the following condition: If $x_1^{j_1}x_2^{j_2}$ is included then
$x_1^{j'_1}x_2^{j'_2}$ are included too for all degrees $j'_1\le j_1$ and
$j'_2\le j_2$.  Otherwise the Legendre polynomials would not have the same
linear span as the monomials.  This condition is satisfied by all practical
background subtraction methods in AFM.

For Gaussian ACF~\eref{G-Gauss-exp} and limited total degree we obtain the
leading terms for small~$\alpha$
\begin{equation}
\beta_j \sim \frac{(j+1)(j+2)}2 \alpha^2
             \left[\pi - \frac{2(2j+3)\sqrt{\pi}}3 \alpha
                   + \frac{j^2+3j+3}3 \alpha^2 \right]\;.
\label{beta-2d-gauss-lo}
\end{equation}
The second term in the brackets must be small compared to 1 for the
leading-order approximation to be valid.
Unfortunately, Gaussian ACF may be the only interesting case for which $\beta$
has a closed form expression because Gaussian is the only separable radially
symmetric function.

\begin{table}
\caption{\label{tab:2d-Cr} Polynomials $C^\mathrm{r}_j$ describing 2D
polynomial background removal in the radially symmetric case.}
\begin{indented}
\item[]\begin{tabular}{@{}llll}
\br
$j$&$C^\mathrm{r}_j(t)$\\
\mr
0&$\pi/2 - 2t + t^2/2$\\
1&$3\pi/2 - 10t+ 7t^2/2 + 8t^3/3 - t^4$\\
2&$3\pi - 28t+ 13t^2 + 56t^3/3 - 9t^4 - 32t^5/5 + 7t^6/3$\\
3&$5\pi - 60t+ 35t^2 + 72t^3 - 43t^4 - 288t^5/5 + 77t^6/3 + 128t^7/7 - 6t^8$\\
4&$15\pi/2- 110t+ 155t^2/2 + 616t^3/3 - 147t^4- 1408t^5/5 + 448t^6/3 + 1408t^7/7$\\
 &${} - 78t^8 - 512t^9/9 + 83t^{10}/5$\\
5&$21\pi/2 - 182t + 301t^2/2 + 1456t^3/3 - 406t^4 - 4992t^5/5 + 616t^6 + 8320t^7/7$\\
 &${} - 534t^8 - 6656t^9/9 + 249t^{10} + 2048t^{11}/11 - 146t^{12}/3$\\
\br
\end{tabular}
\end{indented}
\end{table}

For other radially symmetric ACF $G(t_1L_1,t_2L_2)=G(tL)$, i.e.\ isotropic
roughness, we can obtain leading-order terms using \eref{sigma-C-inf} and
transformation to polar coordinates $t_1=t\cos\omega$ and $t_2=t\sin\omega$.
As in one dimension, this results in an asymptotic series for $\beta$ if
$G(x)$ decays faster than any power $1/x^n$. The integral then becomes
\begin{equation}
\int_0^\infty G(tL)
    \left[ \int_0^{\pi/2} C(t\cos\omega,t\sin\omega) \9\omega \right]
    t \9t
    = \int_0^\infty G(tL) C^\mathrm{r}(t) t \9t\;,
    \label{Gint-infty}
\end{equation}
where the inner integral expressing $C^\mathrm{r}$ is elementary because $C$
is a polynomial.  Polynomials $C^\mathrm{r}$ are listed in \tref{tab:2d-Cr}
for degrees up to 5 for reference.

The outer integral is of the same type as in the 1D case for the same
$G$. For exponential ACF~\eref{G-Gauss-exp} this results in
\begin{equation}
\beta_j \sim (j+1)(j+2) \alpha^2
             \left[\pi - \frac{8(2j+3)}3 \alpha
                   + 2(j^2+3j+3) \alpha^2 \right]\;.
\label{beta-2d-exp-lo}
\end{equation}

\subsection{Intermediate Gaussian--exponential ACF}

Gaussian and exponential ACF belong to a one-parametric class of simple
classical ACF models, usually called power-exponential or intermediate
Gaussian--exponential ACF~\cite{Franceschetti07,Zhao00}.  The parameter is the
power in the exponent:
\begin{equation}
G_p(x) = \sigma^2 \exp[-(x/T)^p]\;.
\label{G-genp}
\end{equation}
Clearly $p=1$ and~2 correspond to exponential and Gaussian~\eref{G-Gauss-exp},
and $p\in[1,2]$ represents a class of randomly rough surfaces transitioning
smoothly between them.
%, as visualized in \fref{fig:gen-gauss}.
The corresponding spectral densities do not have closed forms and the ACF form
is not physically motivated.  Nevertheless, it can match reasonably many real
surfaces.  In this this context the main advantage of this model is that
bias expressions \eref{sigma-C-inf} and \eref{Gint-infty} have simple closed
forms.

In order to derive asymptotic series in powers of $\alpha$ for $\alpha\to0$
(i.e.\ disregarding exponentially small terms), we need to evaluate the
integrals to infinity \eref{sigma-C-inf} and \eref{Gint-infty}.  They have
both the same form
\begin{equation}
\beta = \frac{2^D}{G(0)} \int_0^\infty G(tL) P(t) \9t\;,
\end{equation}
where $P(t)$ is a polynomial--either $C_j(t)$ in 1D (\tref{tab:1d-Cpoly})
or $tC_j^\mathrm{r}(t)$ in 2D (\tref{tab:2d-Cr})---and $G$ is given by
\eref{G-genp}. Writing the polynomial
\begin{equation}
P(t) = \sum_{k=0}^{K} a_k t^k\;,
\end{equation}
we need to evaluate
\begin{equation}
I = 2^D \int_0^\infty \exp\left[-\left(\frac{xL}{T}\right)^p\right]
    \sum_{k=0}^{K} a_k t^k \9t\;,
\end{equation}
which can be easily transformed ($\alpha t=u^{1/p}$) to
\begin{equation}
I = 2^D \sum_{k=0}^{K} a_k \frac{\alpha^{k+1}}{p}
                       \int_0^\infty \rme^{-u} u^{(k+1)/p-1} \9u
  = \frac{2^D}p \sum_{k=0}^{K} a_k \alpha^{k+1}
                               \Gamma\left(\frac{k+1}p\right)\;,
\label{Igenp}
\end{equation}
where $\Gamma$ denotes the gamma function.

Coefficients $a_k$ are given by the basis functions.  For 1D polynomials the
leading coefficients are
\begin{equation}
a_0=j+1\;,
\quad
a_1=(j+1)^2\;,
\quad
a_2=0\;,
\quad\hbox{and}\quad
a_3=j(j+1)^2(j+2)/6\;,
\end{equation}
whereas for 2D polynomials
\begin{eqnarray}
a_0=0\\
a_1=\pi(j+1)(j+2)/4\\
a_2=(j+1)(j+2)(2j+3)/3\\
a_3=(j+1)(j+2)(j^2+3j+3)/12
\end{eqnarray}
Substituting them into~\eref{Igenp} gives leading order terms for 1D bias
\begin{equation}
\beta_j \sim \alpha \frac{2(j+1)}p
             \left[\Gamma\left(\frac1p\right)
                   + (j+1)\alpha\left[\Gamma\left(\frac2p\right)
                                      + \alpha^2 \frac{j(j+2)}6
                                        \Gamma\left(\frac4p\right)
                                        \right]\right]
\end{equation}
and for 2D bias
\begin{equation}
\beta_j \sim \alpha^2 \frac{(j+1)(j+2)}{p}
             \left[\pi \Gamma\left(\frac2p\right)
                   + \alpha \left[\frac{4(2j+3)}3 \Gamma\left(\frac3p\right)
                                          + \alpha \frac{j^2+3j+3}3
                                            \Gamma\left(\frac4p\right)
                                            \right]\right]\;.
\end{equation}
These expressions can be used to reproduce
\eref{beta-1d-gauss-lo} and \eref{beta-2d-gauss-lo} with $p=2$,
\eref{beta-1d-exp-lo} and \eref{beta-2d-exp-lo} with $p=1$,
and similar expressions for ACF of the form \eref{G-genp} for other $p$.

\subsection{Reference plots}

Results of the preceding sections can be summarized in graphical form for
a quick estimation of the bias in common measurement scenarios.  It
always begins with estimating the ratio $\alpha=T/L$, usually by knowing $L$
exactly and estimating~$T$.

\begin{figure}
\includegraphics{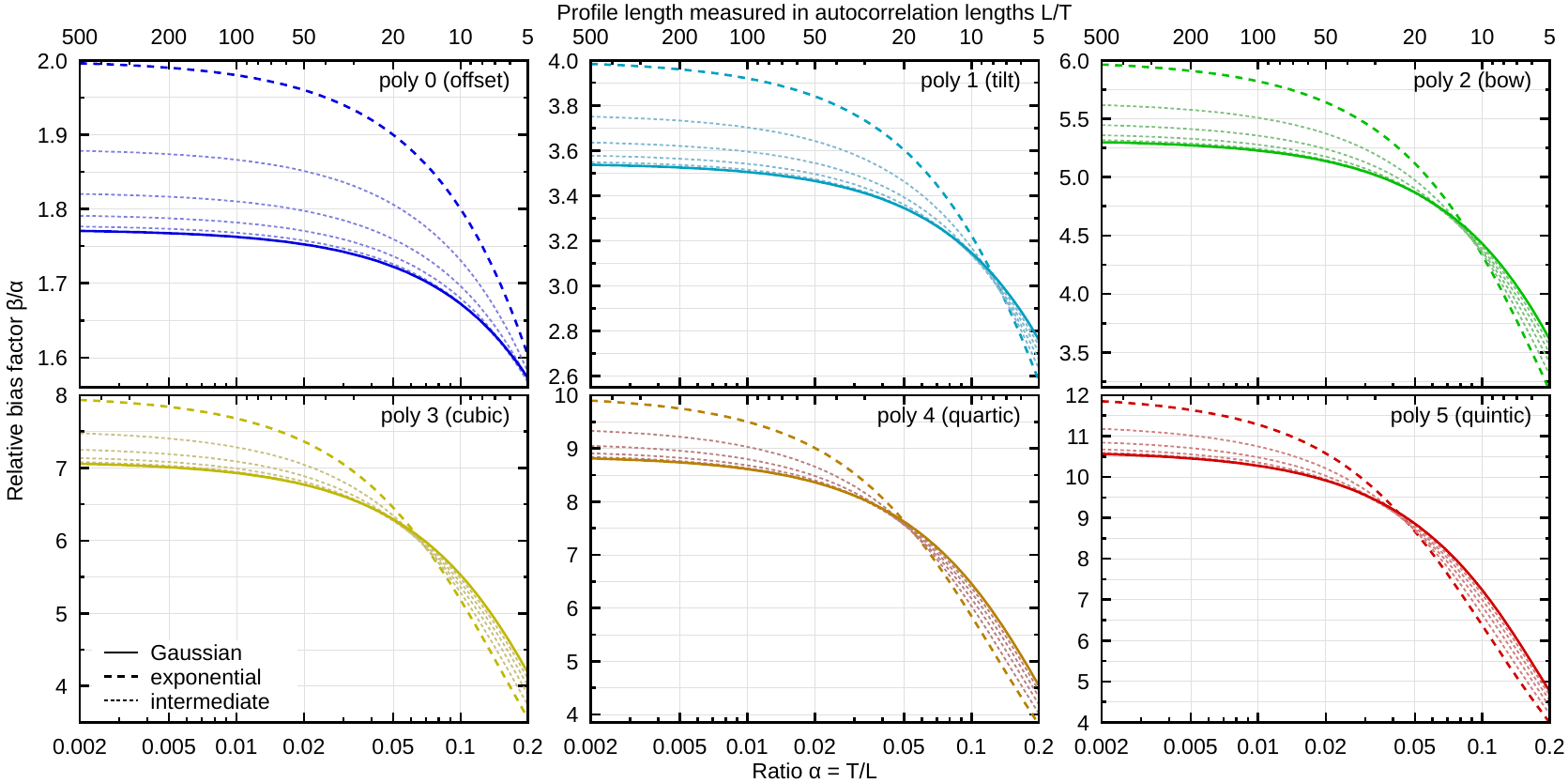}
\caption{Relative bias $\beta$ of $\hat\sigma^2$ for 1D measurements or 2D
measurements with 1D processing (plotted without the leading factor, i.e.\ as
$\beta/\alpha$). Line colour distinguishes the degree of subtracted
polynomial. Line type represents the ACF type---Gaussian ($p=2$), exponential
($p=1$) and intermediate types $p=1.2$, 1.4, 1.6 and~1.8.}
\label{fig:psurvey-1d}
\end{figure}

For 1D processing \fref{fig:psurvey-1d} can be then used.  It summarises the
relative bias $\beta$ for Gaussian and exponential ACF, as well as
intermediate ACF types with $p$ step of 0.2.  It was plotted using exact
integrals and is, therefore, valid even for large $\alpha$.

After choosing the corresponding curve according to polynomial degree and ACF
type, one multiplies the value from \fref{fig:psurvey-1d} by $\alpha$ to
obtain the relative bias $\beta$, and possibly further by $\sigma^2$ for an
absolute number.

An example of bias estimation using \fref{fig:psurvey-1d}:
\begin{enumerate}
\item We measured a $20\times20\rm\,\mu m^2$ AFM image and removed bow from
      each scan line.
\item This means 1D processing, $L=20\rm\,\mu m$ and polynomial degree of~2.
\item We estimate correlation length as $T\approx340\rm\,nm$.  The surface is
      locally smooth and roughness can be assumed not far from Gaussian.
\item This gives $\alpha\approx0.017$, so on the full green curve we
      read~$5.2$.
\item We estimate the bias of $\hat\sigma^2$ as $5.2\times0.017\approx9\,\%$.
\end{enumerate}

The difference between Gaussian and exponential ACF is relatively small in 1D.
The ratio for $\alpha\to0$ corresponds to the ratio $2/\sqrt{\pi}\approx1.13$
of leading terms in \eref{beta-1d-gauss-lo} and \eref{beta-1d-exp-lo}.
However, the difference actually decreases for larger $\alpha$ thanks to the
higher order terms (up to a cross-over point). Furthermore, the curves for
a large range of powers $p$ remain quite close to Gaussian, even up to
$p=1.6$. Assuming a Gaussian ACF can, therefore, often give a reasonable
estimate even if the ACF deviates from Gaussian.

Independently on the ACF type, the bias is quite high.  Even for $L/T$ in the
range of hundreds, it remains at least a few percent and it becomes much
larger as $L/T$ decreases.  It is not difficult to find realistic scenarios in
which it reaches 20, 30 or even 40\,\%.

\begin{figure}
\includegraphics{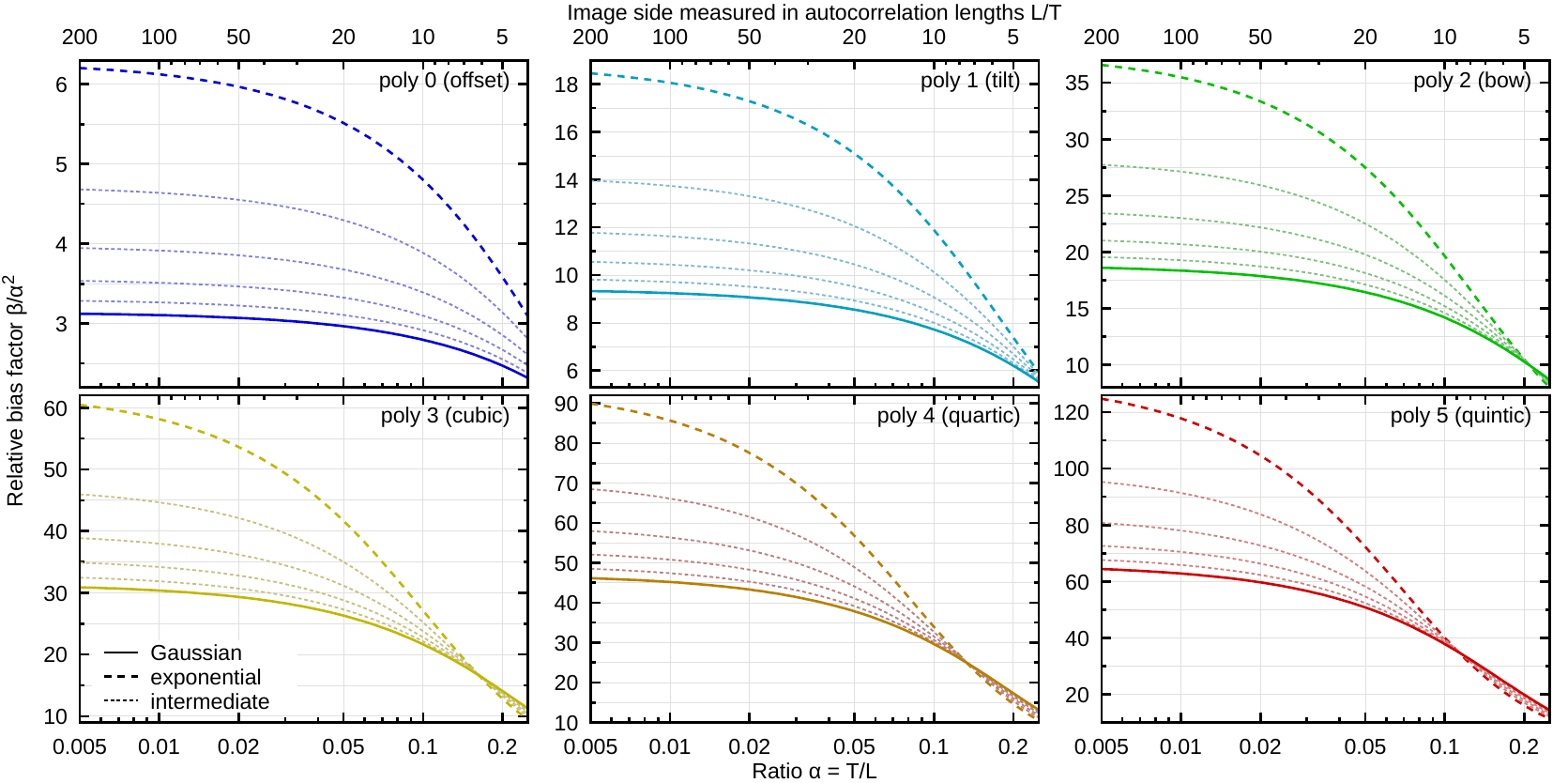}
\caption{Relative bias $\beta$ of $\hat\sigma^2$ for 2D measurements with only
2D background subtraction (plotted without the leading factor, i.e.\ as
$\beta/\alpha^2$). Line colour distinguishes the degree of subtracted
polynomial. Line type represents the ACF type---Gaussian ($p=2$), exponential
($p=1$) and intermediate types $p=1.2$, 1.4, 1.6 and~1.8.}
\label{fig:psurvey-2d}
\end{figure}

For 2D processing \fref{fig:psurvey-2d} can be used instead of
\fref{fig:psurvey-1d}.  Although even for most image data processing the bias
is dominated by 1D processing, a quick check of the 2D levelling contribution
is still useful.  The factors in \fref{fig:psurvey-2d} must be multiplied
by $\alpha^2$ (instead of $\alpha$) since the leading term is proportional to
$\alpha^2$ in 2D.  Otherwise the estimation procedure remain unchanged.  Note
that the polynomial degree in \fref{fig:psurvey-2d} correspond to the limited
total degree.  For other combination of $x$ and $y$ degrees one can utilize
the observation that the leading term is proportional to the number of
coefficients fitted.

The dependence on ACF shape is evidently stronger in 2D than it is in 1D.
The ratio of leading terms is now~2 between exponential and Gaussian ACF.
It still holds that the curves remain closer to Gaussian ACF for relative
large powers $p$.  However, due to larger absolute differences, it is no
longer reasonable to universally assume Gaussian ACF.

Also, the proportionality to $\alpha^2$ means that the bias remains quite low
up to $\alpha$ around 0.05 (depending on polynomial degree).  But once it
becomes non-negligible, it grows rapidly.  Therefore, keeping $L/T$
sufficiently large can be a feasible strategy for avoiding bias caused by 2D
data processing---in contrast to 1D processing.

\subsection{Spatial frequency filtering}

Spatial frequency filtering removes (or suppresses) specific spatial
frequencies.  It is usually done in the frequency space utilizing the Fourier
transform.  It yields the best representation of the data using and sines and
cosines in the least-squares sense
% Bracewell doesn't have it.  Try Marks09 or the Wavelet-sparse-way book.
and thus lies within the same framework.
The basis functions are orthonormal and come in pairs
\begin{equation}
\varphi^\mathrm{cos}_j(x) = \sqrt{\frac2L} \cos\frac{2\pi jx}{L}
\quad\hbox{and}\quad
\varphi^\mathrm{sin}_j(x) = \sqrt{\frac2L} \sin\frac{2\pi jx}{L}\;.
\end{equation}
Therefore, for one particular spatial frequency \eref{C} becomes
\begin{equation}
C_j(t) = \int_0^{L(1-t)} \left[
            \varphi^\mathrm{cos}_j(v) \varphi^\mathrm{cos}_j(v+Lt)
            + \varphi^\mathrm{sin}_j(v) \varphi^\mathrm{sin}_j(v+Lt) \right]
         \9v
\end{equation}
which evaluates to
\begin{equation}
C_j(t) = \frac2L (1-t) \cos(2\pi jt)\;.
\end{equation}
Substituting this $C_j$ expression to \eref{sigma-C} leads to bias
\begin{equation}
\beta_j = \frac4L \int_0^L G_0(x) \left(1-\frac xL\right)
                           \cos\left(2\pi j\frac xL\right) \9x\;,
\label{betaj-freq}
\end{equation}
where $G_0(x)=G(x)/G(0)$ is normalized ACF.  Since ACF is the Fourier
transform of spectral density of spatial frequencies, removing one frequency
from the spectral density corresponds to removing one frequency component from
the ACF.

However, expression~\eref{betaj-freq} is not exactly the $j$-th Fourier
coefficient of ACF.  It would be if $G(x)$ was non-zero only when $x/L\ll1$,
allowing replacing the integral with
\begin{equation}
\frac2L \int_{-L/2}^{L/2} G_0(x) \cos\left(2\pi j\frac xL\right) \9x\;.
\end{equation}
This corresponds to the case $\alpha\to0$.  The difference between
\eref{betaj-freq} and spectral density at frequency $j$ is due to the limited
length and for background removal, i.e.\ small frequencies $j$, it is of order
$\alpha^2$ in 1D.

Since the data spectral density usually has a maximum at the zero frequency
and then monotonically decreases, filtering of low frequencies is the
background removal which most efficiently reduces $\hat\sigma^2$ because it
always takes the largest remaining component.  Nevertheless, for the lowest
frequencies the result is quite similar to the subtraction of polynomials.

% Somewhere here (or earlies?) we might make the following remark.  The
% bias can seem surprisingly large.  It is large because PSD has maximum at
% zero and often then sharply decreases.  So the removal of just a few lowest
% spatial frequencies has large impact.

\subsection{Median levelling}

%\begin{figure}
%\includegraphics{stats-median.pdf}
%\caption{Numerical results for median removal.  Ordinate is the difference
%scaled by the common factors
%$d(\alpha)=E[\hat\delta^2]/(\sigma^2\alpha^D)$
%which approaches a constant for $\alpha\to0$.}
%\label{fig:stats-median}
%\end{figure}

Instead of the mean value, other quantities are sometimes subtracted during
levelling, for instance median or trimmed mean.  The motivation is that they
are less sensitive to outliers.  These operations are non-linear and thus
outside the framework developed above.

For 1D data they are also inconsistent with `mean value of $z$ is zero'.
Furthermore, if we subtracted the mean value afterwards it would nullify the
effect of subtracting something else first.  However, they can be meaningful
for 2D data. When correcting misaligned scan lines, each is levelled
individually using the non-linear operation.  This effect survives subsequent
subtraction of mean value from the entire image---which, in fact, then
frequently has very little effect.

The estimated $\hat\sigma^2$ is again expressed by~\eref{sigma-integral}
if we replace $\hat\mu$ by the subtracted quantity, which will be denoted
$\hat m$ (for median).  It is useful to write $\hat m=\hat\mu+\hat\delta$ as
both $\hat\mu$ and $\hat m$ are location estimates, so their difference
$\hat\delta$ is presumably small.  This gives expected value
\begin{equation}
\E[\hat\sigma^2] = \sigma^2 - \E[\hat\mu^2] + \E[\hat\delta^2]
\end{equation}
as all mixed terms cancel.  Therefore, the negative bias is always slightly
reduced compared to mean value subtraction and the expected difference is
simply $\E[\hat\delta^2]$.  Numerical results confirm this conclusion.
Asymptotic expressions for $\E[\hat\delta^2]$ are known for many distributions
in the case of uncorrelated data.  For instance for median and Gaussian
distribution $\E[\hat\delta^2]=(\pi/2-1)/N$, where $N$ is the number of data
values.  More generally, $N\E[\hat\delta^2]$ tends to a constant for
$N\to\infty$ if the probability density decays sufficiently fast.

For correlated data $N$ again has to be replaced with $\alpha^D$. Numerical
calculations give
\begin{equation}
\E[\hat\delta^2] \approx \sigma^2\alpha^D(p - q\alpha)
\label{median-pq}
\end{equation}
as a reasonable approximations in most cases, with
$p=0.349$ and $q=0.646$ for 1D and Gaussian ACF,
$p=0.293$ and $q=0.436$ for 2D and Gaussian ACF,
and
$p=0.156$ and $q=0.385$ for 2D and exponential ACF.
The exception is exponential ACF in 1D for which
\begin{equation}
\E[\hat\delta^2] \approx \sigma^2\alpha^D p\exp(-q\alpha)
\label{median-pq-exp1d}
\end{equation}
with
$p=0.182$ and $q=3.34$ is more suitable for covering a wider $\alpha$ range.
In both formulae $p$ corresponds to the limit $\alpha\to0$ and we can put
$q=0$ for a rough estimate.

Considering the small differences between biases for mean and median
levelling, detailed analysis of trimmed means is unnecessary.  The bias lies
between values for mean and median---and this is sufficient for its
estimation.

%{\it XXX For median of differences (or more generally trimmed mean of
%differences), we should do some numerical simulations first.  It is a nice
%method in practice but difficult to analyse.}

\section{Bias of other quantities}
\label{sect:other-bias}

So far, we only considered the bias of $\hat\sigma^2$.  It has a linear
definition, making it suitable for averaging, and it often arises naturally in
physical calculations, for instance in optics. Replacing
$\hat\sigma^2$ with $\hat\sigma^2/(1-\beta)$ corrects its bias. However, there
are many other quantities characterizing the extent or variance of heights of
rough surfaces.  We will consider unsquared $\sigma$ and average
roughness, denoted Ra (whether in 1D or 2D).  Finally, we will introduce
a single symbol for $\sigma^2$ to avoid confusing notation: $s=\sigma^2$.

It might seem that if we correct $\hat s$ by dividing by $1-\beta$ then
$\hat\sigma$ corrected should simply use $\sqrt{1-\beta}$.
Unfortunately, this is only true in the limit $\alpha\to0$.  For finite
$\alpha$ correcting by $1/\sqrt{1-\beta}$ does not result in an unbiased
estimate.  The reason is that $\hat s$ has non-zero dispersion (proportional
to $\alpha^{D/2}$~\cite{Zhao00}) and square root is a non-linear
transformation.  Since square root is concave, the Jensen's
inequality~\cite{Jensen06,Krishnan15} states that $\sqrt{\hat s}$
underestimates $\sigma$ when $\hat s$ itself is unbiased.

The Taylor expansion of $\sqrt{s}$ around $\E[s]$ gives an expression in term
of $n$-th central moments $\mu_n[\hat s]$~\cite{Kendall77}
% This should also be in Krishnan15?
\begin{equation}
\E[\hat\sigma] = \sigma \left[1
                              - \frac18\frac{\mu_2[\hat s]}{s^2}
                              + \frac1{16}\frac{\mu_3[\hat s]}{s^3}
                              - \frac{15}{128}\frac{\mu_4[\hat s]}{s^4}
                              + \dots \right]\;.
\label{Esigma}
\end{equation}
Although the values entering \eref{sigma-expanded} are correlated, the law of
large numbers still means $\hat s$ will tend to the normal distribution (the
dispersion of heights $z$ is obviously finite).  Therefore, we can estimate
$\mu_{2k+1}[\hat s]\approx0$ and
$\mu_{2k}[\hat s]\approx \mathrm{Var}[\hat s]^k(2k-1)!!$, which hold for
central moments of the normal distribution.
% Krishnan15 and most probability books contain these.
However, the variance
$\mathrm{Var}[\hat s]$ depends on the dimension, ACF type, levelling method
and is, of course, a function of $\alpha$.  The leading term is~\cite{Zhao00}
\begin{equation}
\mathrm{Var}[\hat s] \sim a s^2 \alpha^D\;,
\label{sigma2-variance}
\end{equation}
where $a$ is a constant for given ACF.  In general, \eref{sigma2-variance} is
again a series in $\alpha$.  Together with \eref{Esigma}, this again give a
series expression
\begin{equation}
\E[\hat\sigma] = \sigma \bigl(1 - a_1\alpha^D - a_2\alpha^{2D} - \dots\bigr)
\end{equation}
for the biased mean value of $\hat\sigma$.  The bias is again negative and can
be corrected by dividing by the term in parentheses.

One consequence of relation \eref{sigma2-variance} which needs to be
emphasized is that there is a difference between averaging $M$ independent
profiles and $M$ correlated image rows.  The bias $\beta$ is the same in both
cases if row-wise processing is applied.  However, the variance of $\hat s$
is reduced by factor $1/M$ for independent scan lines, whereas for the image
it is reduced only by $\alpha=T/\Delta_y\times1/M$, where $\Delta_y$ is the
vertical sampling step.  So both the variance and the bias originating from
\eref{sigma2-variance} are larger for correlated image rows.

\begin{figure}
\includegraphics[width=8.5cm]{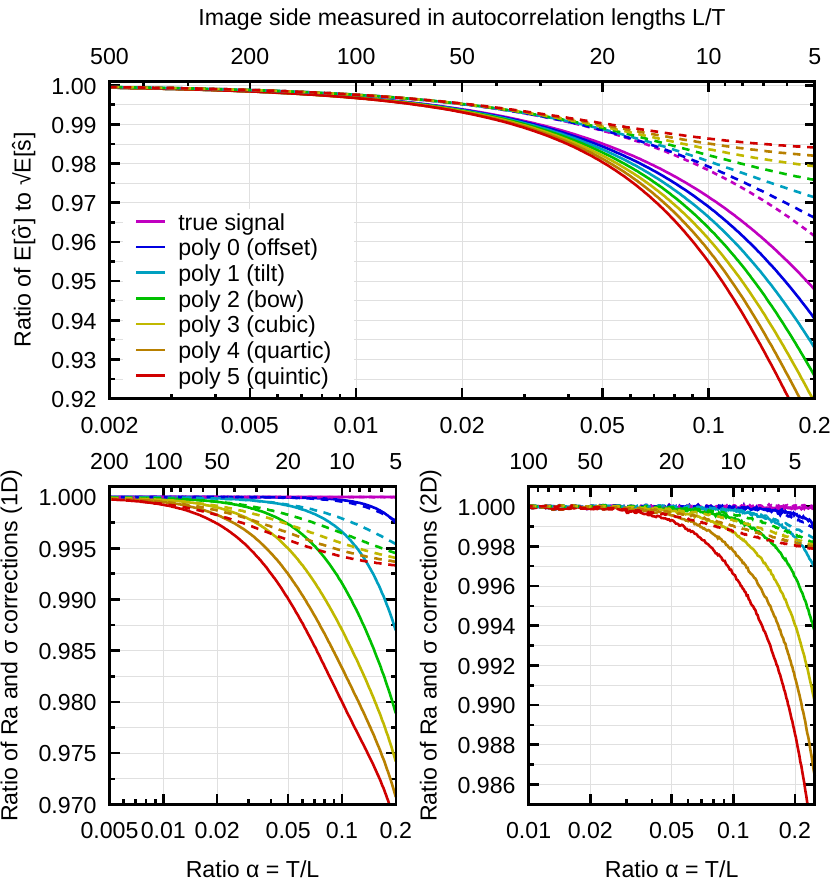}
\caption{Results of simulation showing the bias due to non-linearity bias
not captured in error propagation rule (top) and ratio of correction factors
for Ra and $\sigma$.  Dashed lines correspond to
exponential ACF, full lines to Gaussian.}
\label{fig:stats-other}
\end{figure}

Nevertheless, the bias following from variance is largest for single profiles,
where no averaging reduces it.  Its magnitude is illustrated in
\fref{fig:stats-other} for this 1D case and Gaussian and exponential ACF.
Even in this case it does not exceed 1--2\,\% for reasonable $L/T$ ratios,
although it is somewhat higher for the Gaussian ACF.  For other cases the bias
is negligible since it is smaller by at least another order of magnitude.

Concerning Ra, the ratio Ra/$\sigma$ is a constant for any particular
distribution of heights. Therefore, in the limit $\alpha\to0$ the correction
factor for Ra---or any other quantity characterizing linearly the variance of
heights---is the same as for $\sigma$.  Of course, the main point of this work
is that $\alpha$ cannot be considered zero.  The distribution of heights
changes somewhat by levelling, so we must ask how much the correction factors
change with increasing $\alpha$.

The results of numerical calculations are plotted in \fref{fig:stats-other}.
Fortunately, the ratios of correction factors for Rq and Ra are close to
unity, even though they are much larger for Gaussian ACF than for exponential.
In 2D the effect can be probably safely disregarded, in 1D it may be useful to
consider it, depending on the ACF form.

\section{Experimental example}

\begin{figure}
\includegraphics{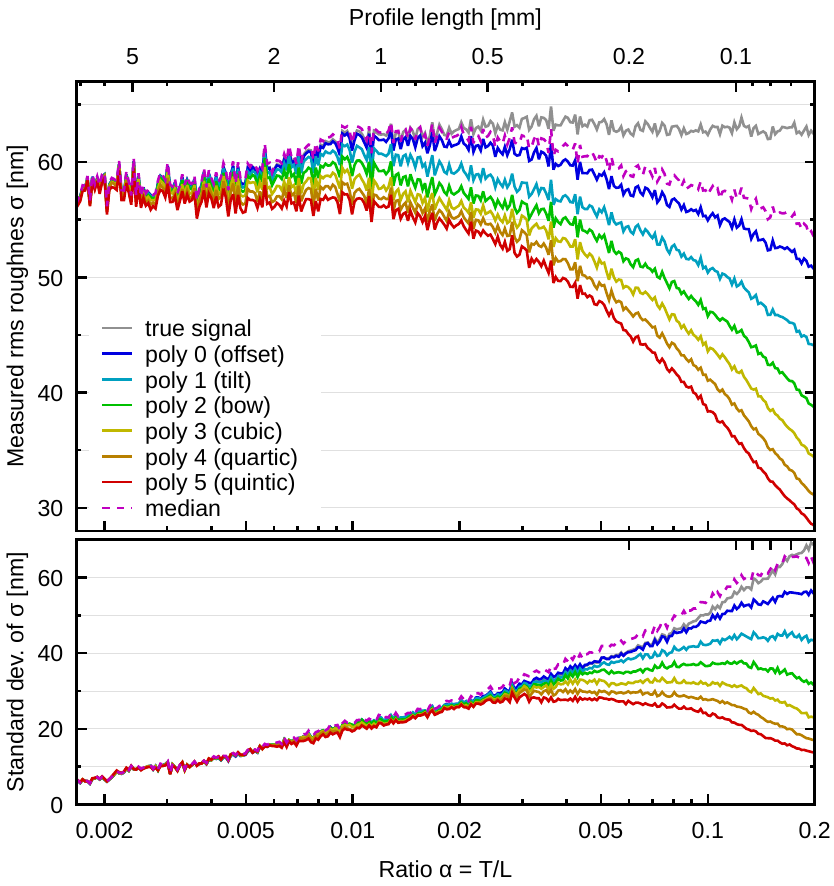}
\caption{Measured roughness and its standard deviation as a function
of profile length for a reference roughness sample with Ra of 50\,nm, showing
the bias evolution for shorter profile lengths and its dependence of the
levelling.}
\label{fig:stats-NMM-Ra0.05um}
\end{figure}

A realistic example illustrating the impact of profile length on measured
roughness quantities is shown in \ref{fig:stats-NMM-Ra0.05um}.  It was
obtained by measuring a set of long profiles of a surface roughness standard
from Edmund Optics based on electroformed nickel plates representing different
surface finishes.
Measurements were done using the Nanomeasuring and Nanopositioning
Machine NMM1~\cite{Hausotte07} from SIOS company.  Combined with a custom
built AFM head (used in contact mode here with PPP-CONTR cantilevers), the
instrument can be used for measurements over even a centimetre
areas. The measured profiles were approximately 1.2\,mm long,
approximately 1000$\times$ longer than the estimated correlation length of
12$\,\rm\mu m$. Data corresponding to shorter evaluation lengths were then
obtained by cutting short segments from these profiles.

The dependency of measured mean square roughness
$\sigma$ on $\alpha=T/L$ is plotted in \fref{fig:stats-NMM-Ra0.05um}
for each polynomial degree from 0 to 5 and for illustration for median
levelling as well (even though it does not satisfy the zero-mean assumption).
Overall, the dependencies resemble the theoretical curves, as illustrated
for instance in \fref{fig:stats-1d-gauss} for Gaussian ACF.  The decrease of
$\sigma$ for smallest $\alpha$ is an artefact caused by levelling of the long
base profile.  The longer profiles cut from it were already of comparable
lengths; a longer base profile would be necessary for stable result.

\Fref{fig:stats-NMM-Ra0.05um} also illustrates the standard deviation of
measured $\sigma$ as a function of $\alpha$.  According to the asymptotic
estimates~\cite{Zhao00} it should not depend on the levelling for small
$\alpha$.  This is confirmed as up to $\alpha\approx0.02$ the curves
are indistinguishable.  Considering contributions to measurement uncertainty,
this random part predominates, at least for a single evaluation.  However, it
can be reduced by evaluating roughness multiple times, which is anyway
recommended.  In contrast, the bias is unaffected by repeated measurement
because it is inherently tied to~$T/L$.

\section{Conclusion}

Bias caused by limited measurement area is a universal and mostly unavoidable
effect skewing measured roughness values.  While the effect itself is well
known (mostly for the case of mean value subtraction), it is not commonly
taken into account, neither in roughness measurement standards, nor in
practice.  It is made worse by the use of aggressive levelling in atomic force
microscopy data processing as the subtraction of higher order polynomials
increases the bias.  And it is further exacerbated by levelling of
topographical images scan line by scan line.  While this step is often
necessary, it turns 2D data processing to 1D, at least from the bias
standpoint---we are then effectively analysing profiles of the length of one
scan line.

We developed a framework for the calculation of the bias using known data
autocorrelation function (ACF) for levelling by subtraction of any
linearly-fitted background.  Beside exact results, it allowed providing
expressions in the form of series for 1D and 2D data polynomial levelling and
common ACF types.  We noted that in 1D processing, the dependency on ACF type
is relatively weak and it seems to be possible to simply assume Gaussian ACF
if the surface is close to locally smooth. Some common levelling methods are
non-linear.  Of these, we examined median levelling and found that it is
similar to mean value subtraction as its effect is essentially the removal of
one degree of freedom. Finally, the translation of results for squared mean
square roughness to other quantities was discussed.

For an easy rough estimation of the bias, two sets of reference plots were
provided for both 1D and 2D data processing (figures \ref{fig:psurvey-1d}
and~\ref{fig:psurvey-2d}), covering a range of ACFs from Gaussian to
exponential, including several intermediate types, and levelling polynomial
degrees from 0 to~5.  After estimating the ratio $\alpha$ of correlation
length to scan line length, they allow obtaining the \emph{relative} negative
bias $\beta$ of $\hat\sigma^2$.  For Ra, Sa, Rq or Sq the bias is
approximately half this value.

%{\it TODO some final words...}
%Always measure more than one profile or image.

\subsection*{Acknowledgements}

This work was supported by the EURAMET joint research project "Six degrees of freedom" funded from the European Union's Seventh Framework Programme,
ERA-NET Plus, under Grant Agreement No. 217257, and
by the Ministry of Education, Youth and Sports of the Czech Republic under the
project CEITEC 2020 (LQ1601).

\subsection*{References}
\bibliographystyle{unsrt}
\bibliography{references}

\end{document}